\documentclass[12pt]{article}
\usepackage{graphicx}
\usepackage{color}
\usepackage{amsmath,slashed, amssymb}
\usepackage{fancyhdr}
\usepackage{hyperref}
\usepackage[titletoc]{appendix}
\numberwithin{equation}{section} 

\def\beq{\begin{eqnarray}}
\def\eeq{\end{eqnarray}}
\def\bea{\begin{eqnarray*}}
\def\eea{\end{eqnarray*}}

\def\centeron#1#2{{\setbox0=\hbox{#1}\setbox1=\hbox{#2}\ifdim
\wd1>\wd0\kern.5\wd1\kern-.5\wd0\fi
\copy0\kern-.5\wd0\kern-.5\wd1\copy1\ifdim\wd0>\wd1
\kern.5\wd0\kern-.5\wd1\fi}}
\def\ltap{\;\centeron{\raise.35ex\hbox{$<$}}{\lower.65ex\hbox{$\sim$}}\;}
\def\gtap{\;\centeron{\raise.35ex\hbox{$>$}}{\lower.65ex\hbox{$\sim$}}\;}

\newcommand{\newc}{\newcommand}
\newc{\qbar}{{\overline q}}
\newc{\Kahler}{Kahler }
\newc{\deltaGS}{\delta_{\rm GS}}
\begin{document}
\begin{titlepage}
\begin{flushright}
{\large SCIPP 15/15\\
}
\end{flushright}

\vskip 1.2cm

\begin{center}

{\LARGE\bf Classical and Quantum Stability in Putative Landscapes}

\vskip 1.4cm

{\large Michael Dine$^{(a)}$
}
\\
\vskip 0.4cm
{\it $^{(a)}$Santa Cruz Institute for Particle Physics and
\\ Department of Physics, University of California at Santa Cruz \\
     Santa Cruz CA 95064  } \\
\vspace{0.3cm}

\end{center}

\vskip 4pt

\vskip 1.5cm

\begin{abstract}
Landscape analyses often assume the existence of large numbers of fields, $N$, with all of the many couplings among
these fields (subject to constraints such as local supersymmetry)
selected independently and randomly from simple (say Gaussian) distributions.  We point out that unitarity and perturbativity place significant constraints on behavior of couplings with $N$, eliminating otherwise puzzling results.  In would-be flux compactifications
of string theory, we point out that in order that there be large numbers of light fields, the compactification radii must scale as a positive power of $N$; scaling of couplings with $N$ may also be necessary for perturbativity.  We show that in some simple string
theory settings with large numbers of fields, for fixed $R$ and string coupling, one can bound certain sums of squares
of couplings by order one numbers. This may argue for strong 
correlations, possibly calling into question the assumption of uncorrelated distributions.
We consider implications of these considerations for classical and quantum
stability of states without supersymmetry, with low energy supersymmetry arising from tuning of parameters,
and with dynamical breaking of supersymmetry.
\end{abstract}

\end{titlepage}

\section{Introduction}

The notion that there is an underlying landscape leads to a picture in which there is an ensemble of theories with varying types and number of degrees of freedom at a high energy scale, and with lagrangian parameters at that scale picked at random from an underlying distribution.  Flux vacua in various theories have served as models for understanding these questions\cite{boussopolchinski}.  If
there are large numbers of quantized fluxes, each of which can take many
values, the number of states can be exponentially large.  Many analyses of this possibility have been conducted in string theory (e.g. IIB and F theory).  In these theories, if the fluxes
vanish, there are  large numbers of moduli; moduli and fluxes are paired.  Many of these fields may be stabilized and gain mass
once fluxes are turned on, already at the level of the classical equations.  The stabilization of these moduli is usually studied at the level of model effective field theories.  In many cases\cite{douglasdenef1,douglasdenef2} the working assumption has been that one could describe the system with a locally supersymmetric lagrangian with a large number of fields, with Kahler potential and superpotential terms chosen from suitable (say Gaussian) distributions.  Many of the results obtained from these studies appear robust, with a generic flavor.  For example, distributions of supersymmetry breaking scales for small scales can be understood in terms of a low energy theory of a single chiral field, with parameters chosen from uniform distributions\cite{dinesun}.

Another approach to modeling a landscape is to simply consider a field theory with a large number of fields, and make assumptions about the statistical distributions of lagrangian parameters\cite{nimasavasshamit,easther}.%,susskindmanyfields}.
We will refer to this as the {\it large N} viewpoint.  The relation noted above, between numbers of possible fluxes in string compactifications and moduli, suggests to many authors a connection between these two pictures.  Much of our discussion in this paper will be in the context of systems with large numbers of fields.  In such theories, it is relatively easy to formulate the
problem of finding
stationary points, and to consider issues of classical and quantum stability.

On the other hand, in a scenario in which most or all moduli are stabilized
by a potential generated by fluxes, moduli masses typically scale as an inverse power of the compatification radius (for simplicity, we will think of a compatification with all compact dimensions of comparable size, $R$).  In
this case a large number of light pseudomoduli requires that $R$ be large; if this is not the case, it is not clear why the analysis in terms of many
fields is sensible.  We will
argue that large $R$ is not likely to play a special role in determining the statistics of vacuum states, and that a priori one expects such states to be rare.  In flux compactifications, then, reconciling
large $N$ results, such as the fraction of stationary points exhibiting stability, with this expectation will provide consistency
checks on assumptions about the underlying landscape.

In would-be vacua without supersymmetry, the questions of classical stability\cite{easther,mcalister} and quantum metastability\cite{greeneweinberg,dinepaban}
have largely been explored in the large N framework (exceptions, for quantum tunneling, include\cite{dinefestucciamorisse}). 
These analyses make various assumptions about the low energy field content
and statistics of the effective action of such states.  An often unstated assumption must be made from the start:  the couplings at these
stationary points must be sufficiently weak that a semiclassical analysis is sensible; otherwise the notion of ``stationary points" is not meaningful.  Accepting this (and the conditions for and plausibility of this assumption will be one of the topics of this paper), naively, one might expect that among non-supersymmetric stationary points, a fraction of order $(1/2)^N$ would be stable or metastable.  Various studies\cite{easther,mcalister}, however, suggest far more severe suppression, possibly as $e^{-cN^2}$ or more generally $e^{-aN^p}$, for some constant positive constants, $a,~c$ and $p$. 
For approximately supersymmetric states, a high degree of metastability would seem almost assured\cite{dinefestucciamorisse},
but the results of \cite{mcalister,Bachlechner:2014rqa} suggest that one should rethink this issue as well.
%With approximate supersymmetry (we will make this notion more precise later), some studies\cite{mcalister,Bachlechner:2014rqa} find $e^{-cN}$.  This latter result is somewhat puzzling, as one expects only a small number of light chiral fields to play a role in supersymmetry breaking. 
As for quantum stability, among metastable states, without supersymmetry, one finds an exponential
suppression of stability against tunneling, but the analyses make a number of assumptions, and it is not clear how general they are.

In this note we attempt to develop a more robust understanding of the question of stability. 
%We first consider, in the next section, a simple class of models with multiple fields which illustrates the main issues.
In section \ref{nfieldmodels}, we discuss some general aspects of $N$-field models.
In section \ref{nscaling},
under the assumption that couplings of fields are uncorrelated, we determine how couplings must scale with $N$
in order that there be a sensible perturbative expansion in the low energy effective field theory; more precisely that 
the low energy theory should not be in violent conflict with unitarity.   In section \ref{stringmodels}, we consider implications
of these scalings for actual string models.  We first note that in flux compactifications, the existence of large numbers of
relatively light fields requires that the compactification radius grow as a power of the coupling, and discuss the power
with particular frameworks.  We argue that it is not plausible,
with or without supersymmetry, that the preponderance of metastable states lie at large radius, and consider the connection of this issue with the
question of stability for large $N$.
We also note the possibility that, independent of the coupling and radii, sums of (squared)
couplings do not grow with $N$.  This question can be investigated in critical string theories, and in 
section \ref{stringcalculation}, we carry through a simple computational test, demonstrating that indeed certain sums of squared couplings are of order one rather than order $N$.   We argue that this is an indication of strong correlations among couplings.%, rather than simply a uniform scaling of couplings.

In section \ref{approximatesusy}, we consider states with approximate supersymmetry, with a particular focus on questions
of stability and the possible role of large numbers of fields.  We first review two models for supersymmetry in a landscape, one in which the hierarchy is achieved by tuning (of fluxes), another where it arises due to dynamical
breaking of supersymmetry (in some form).  We explain why one expects
that stability, both classical and quantum, in a landscape is a feature of an order
one fraction of stationary points.  In the tuned case, we review the result of \cite{mcalister,Bachlechner:2014rqa}, that stability
is suppressed exponentially in the number of fields, $N$.  We explain why this is surprising, and isolate the origin of the suppression.
We argue that,
if the requirements for unitarity (perturbativity) are satisfied, classical stability is typical. % and note that it results from the assumption that a large number of couplings of a particular type are chosen independently and at random from a statistical distribution.  We point out, first, that this assumption violates unitarity, parameterically, by a power of $N$, without strong assumptions about the distributions.  \footnote{This unitarity violation would correspond to a breakdown of perturbation theory in string theory for surprisingly small coupling.}  Alternatively, if unitarity is respected, there is no exponential suppression.
%In section \ref{nfieldsfluxmodels}, we consider the extent to which flux vacua, for fixed values of the flux, might be well modeled by theories with large numbers of fields.
In section \ref{approximatesusy}, we review ideas for different branches of the 
landscape, and discuss the implications of our observations.  With small supersymmetry breaking, either through tuning or
(especially for the case of) dynamical supersymmetry breaking, stability is not a significant constraint.     In section \ref{nonsusy}, we conclude, with an overview and some final speculations on the possible role of supersymmetric and non-supersymmetric states in a landscape.
% speculate on the behavior of putative branches of the landscape without supersymmetry.  We point out that if couplings do not
%scale uniformly with $N$, some of the arguments for severe suppression of metastability do not hold.

\section{Multiple Field Landscape Models}
\label{nfieldmodels}

One picture for the emergence of a discretuum or landscape of states is that of Bousso and Polchinski\cite{boussopolchinski}, in which there are a large number of possible choices of fluxes, each defining a system with
a unique (or small number of) (metastable) ground state(s).  Another is that there are large numbers of fields, $N$, admitting a vast number of stationary points, some fraction of them stable\cite{nimasavasshamit}.  In Calabi-Yau compactifications of string theories, there would appear to be a connection.  In the case of IIB theories, for example, the number of possible $(2,1)$ forms is related to the number of complex structure moduli, which are massless in the absence of fluxes.  So if one number is large, the other is large.

Theories with many fields provide a relatively simple setting in which to address questions of classical and quantum stability.
As a simple starting point, which illustrates a number of the main issues, consider a theory with $N$ scalars, where the potential is simply a sum of terms
\beq
V(\vec \phi) = \sum_{i=1}^N f_i(\phi_i)
\eeq
Assume that the $f_i$'s are all bounded below, and grow without bound at infinity.  If a typical $f_i$ has $n$ minima and $n-1$ maxima, then the system has $(2n-1)^N$ stationary points and $n^{N}$ minima.  In other words, an exponentially
small fraction of stationary points, ($2^{-N}$ if $n$ is large), are classically stable.   This picture may change drastically,
as discussed in \cite{easther}, once couplings of the different fields to each
other  are considered.  If all couplings between fields are of order the couplings in the $f_i$'s, an assumption we will refer to as {\it coupling democracy}, we might expect significantly more stationary points.  At the same time, only an extremely tiny fraction would be true minima.  This point was
stressed in \cite{easther},
The suppression of stability, these authors found, is of order $e^{-cN^{2}}$, for some constant $c$.  If the couplings scale, instead, as some inverse power of $N$, or if only a few couplings are typically appreciable, then, without supersymmetry, the suppression
of stability is only exponential. %Much of our focus in this paper will be on the plausibility of this democratic assumption.

Quantum stability can be addressed in a similar way.  The authors of \cite{greeneweinberg} considered tunneling in a theory with many fields.  Assuming the existence of a minimum, they took the coefficients of the Taylor series expansion about that point to be 
independent random numbers.  In numerical simulations, they found an exponential suppression of states with (logs of) lifetimes   
greater than some fixed value.  These results were understood analytically in \cite{dinepaban}.  The analysis was particularly simple in the case where couplings between different fields were suppressed.

Ref. \cite{mcallister} considered the question of classical stability in the framework of locally supersymmetric
lagrangians.  These authors selected Kahler potentials and superpotentials from plausible random distributions. Without approximate supersymmetry, they found a suppression with $N$ behaving as $e^{-c N^{1.3}}$.  We will discuss their
results in the case of approximate supersymmetry shortly.
%Supersymmetry alters these considerations.  Here (in the context of local supersymmetry) we might consider a theory with polynomial superpotential and Kahler potential whose coefficients are random numbers.  An interesting set of results obtained from such analyses are those of \cite{mcallister}.  These authors considered the question of classical stability.  Considering a supergravity effective action, they asked for what fraction of
%stationary points are all eigenvalues positive.  Naively, with $N$ fields, one might expect a suppression by a factor $(1/2)^N$. For non-supersymmetric stationary points they found a much stronger suppression, of order $e^{-cN^2}$, for some constant $c$.  %Among the subset of states exhibiting approximate supersymmetry \cite{mcallister} and \cite{Bachlechner:2014rqa} found a suppression $e^{-a N}$.
One might argue that the use of a supersymmetric effective lagrangian is a strong assumption.  Unless the scale of supersymmetry breaking is small, it is not clear why one should cut off the derivative expansion for the effective field theory at terms with two derivatives.  This corresponds to cutting off the theory at terms with at most two powers of auxiliary fields.   At least in cases where the supersymmetry breaking scale does turn out to be low, however, one can argue that the analysis is self-consistent. This suggests that the truth might lie in between the results of \cite{easther} and \cite{mcallister}.  In either case, this is a very substantial suppression, suggesting that metastable states, at least without some approximate supersymmetry,
might be quite rare.  We will comment more on this issue shortly.

\section{Scaling with $N$}
\label{nscaling}

In a theory with $N$ fields, the assumption that all couplings in an effective lagrangian
are selected at random, from $N$-independent distributions, is not sensible as
$N$ gets large.  The theory will violate unitarity
in perturbation theory. Radiative corrections will grow with powers of $N$ (in the supersymmetric case, we will
discuss a potential violation of unitarity already at tree level shortly).  
Indeed this violent breakdown of
perturbation theory will indicate a lack of even a qualitative understanding of the physics.

If we demand that couplings scale uniformly with $N$ so as to yield a sensible expansion, and assume, for simplicity, that the scale
of non-renormalizable couplings is comparable to the ultraviolet cutoff (an assumption which we will assess later in
the context of string theory), we can, by considering loop corrections, obtain scaling laws as a function of $N$.  In particular, if we insist that successive terms in the perturbation expansion be as small or smaller by powers of $N$, and if there is just one characteristic
mass scale, which we take , for simplicity, to be the Planck scale, $M_p$, we have
\begin{enumerate}
\item  Without supersymmetry, terms in the potential of the form $\phi^{a}$ (we will confine our attention to $a>2$) should have couplings which scale with inverse
powers of $N$.  Writing these as
\beq
\gamma_{ijk\ell...} \phi^i \phi^j \phi^k \phi^\ell \dots
\eeq 
then the scaling
\beq{\gamma^{(a)} \sim M_p^{-(a-4)}}~{ N^{-{a\over 2}}}.
\label{gammascaling}
\eeq
yields systematically small perturbative corrections.
To see this, it is helpful to first examine some examples.  Consider corrections to the two point function.  If we consider $<\phi_i \phi_i>$ ($i$ not summed), then there is a correction, quadratic in the four point couplings, proportional to $N^3$.  With the scaling above, the result behaves as
$N^{-1}$.    If we have different indices, $<\phi_i \phi_j>$, we might expect the different contributions to add with random signs, and the result is
further suppressed.  Similarly, for the contribution from the six point, eight point functions.  Corrections to the four point function are similarly
suppressed.

One can possibly allow a slightly weaker suppression with $N$.  The $a$-point functions which receive the largest corrections are those with
some indices identical.  It is possible that these are larger, by powers of $N$, than those with all indices different.  In particular, the scaling for the couplings with different values for the indices,
\beq
{\gamma^{(a)} \sim M_p^{-(a-4)}}~{ N^{{1-a\over 2}}}.
\label{gammascalingprime}
\eeq
while for those with indices equal in pairs:
\beq{\gamma^{(a)} \sim M_p^{-(a-4)}}~{ N^{{2-a\over 2}}}.
\label{gammascalingequalindices}
\eeq
appears sensible.   We will not pursue this further, adopting the stronger scaling if necessary.

%  For the four point function, there is a correction to $\gamma^{(4)}$, at second order.  There are $N^2$ particles in the loop (at one loop).  If the initial state particles, labeled by $ij$,
 % and and final state particles, labeled by $kl$ are different, and assuming their
%contributions combine with random phases, the correction has the structure
%\beq
%\gamma_{ijmn} \gamma_{mnk\ell} \propto \gamma^4 N.
%\eeq
%This yields the scaling law above, for the case $a=4$.  However, if $k \ell = ij$, the contributions all combine with the same sign,
%and we have:
%\beq
%\delta \tilde \gamma_{ijij} \propto \gamma_{ijmn}^2 \propto N^0.
%\eeq
%So these ``diagonal couplings" are intrinsically larger than the off-diagonal couplings.  Their insertion in loops, however, does not lead to an enhancement.  We won't give a complete generalization here, but note, for example, that for a coupling of $2n$ fields, if $n$ pairs of indices are the same, the coupling behaves as 
%\beq
%\tilde \gamma^{(2n)} \propto N^n N^{-n/2 \times 2} = N^0.
%\eeq

\item  With supersymmetry, if the couplings  in the superpotential with $a$ powers of chiral fields scale as
\beq
\lambda^{(a)} \sim M_p^{3-a}~N^{-{(a-1) / 2}}.
\label{nscalings}
\eeq
\end{enumerate}
then perturbation theory appears under control.  For example, the corrections to the Kahler potential scale as $N^0$.
One can check that this coincides with a somewhat more severe suppression
of scalar couplings in \ref{gammascaling}.  The quartic coupling, for example, scales as $N^{-2}$.   It should be stressed, again, that we are assuming here that there is just one relevant mass scale in the underlying theory, and that, apart from the scalings with $N$, there are no other parameterically very small parameters.  Such parameters (small pure numbers or ratios of mass scales) could alter the considerations described here.

%In the non-supersymmetric case, with real fields, these scalings are readily seen by considering two point functions of the form $\langle \phi_i \phi_i \rangle$.
%At one loop, there are $N^{n-1}$ same-sign contributions.  At higher orders, there are more contributions with random signs; the potentially most problematic scale in the same way.  Higher $n$-point functions behave similarly.  In the supersymmetric case, these scalings can be seen
%quickly by considering two point functions of the form $\langle \phi_i^\dagger \phi_i \rangle$, where the $\phi$'s are chiral fields.  The counting is
%as above.  One can check that these scalings lead to the same scalings for the component fields as in the non-supersymmetric case; in making
%this comparison, it is important to note that, for example, the quartic couplings arise as sums of $N$ terms with, in general, random phases.

We can compare these behaviors with what
one would expect if, in some basis for the fields, only the self couplings were of order
one, while all of the others vanished (as some large power of $N$).  We will refer to such couplings
as {\it sparse}.  For the case of real fields, for example,
one could consider transforming the fields to another basis by a random $O(N)$ matrix, leading to couplings scaling as inverse powers of $N$.  Then, for example, for the four point coupling,
\beq
\gamma_{ijkl} = \sum_m O_{im} O_{jm} O_{km} O_{\ell m} \tilde \gamma_{mmmm}.
\eeq
Typical entries in the orthogonal matrix $O$ would behave as $N^{-1/2}$, and assuming terms add with random signs, we would
have
\beq
\gamma_{ijkl} \propto N^{-3/2}.
\eeq
More generally, for a coupling involving $a$ real scalar fields, 
\beq
\gamma^{(a)} \propto N^{1 -a \over 2}.
\eeq
This corresponds to our weaker condition, \ref{gammascalingprime}.  For couplings with all indices paired, again we find
a similar scaling as in equation \ref{gammascalingequalindices}.

\section{Lessons From String Theory}
\label{stringmodels}

Modeling of landscapes has been heavily influenced by string theory; indeed, this is the only framework we have in which to make sense of higher dimensional theories, or even four dimensional theories with gravity.  As we have stressed from the
beginning, fluxes in string theory have provided both a key to how exponentially large numbers of states might
arise, and to how large numbers of relatively light scalar fields might emerge.  On the other hand, if the candidate
light fields are the moduli which exist in the absence of fluxes, and these fields gain mass proportional to flux,
it is not clear why, in general, these fields will be light compared to other string states.  The standard (and most plausible)
suggestion is that this will occur when compactification radii are large.

\subsection{Conditions on Radii and Couplings}

In the IIB theory, for example, complex structure moduli acquire potentials in the presence of fluxes.   If typical fluxes are of order $M$,
there is a $2N\times 2 N$ mass matrix for the scalars, with typical entries
\beq
\left ( m_{csm}^2 \right )_{ij}  \propto {M^2 \over R^6}
\eeq
(possibly with additional powers of $N$).  If the matrix is very sparse, the eigenvalues are of order $M^2$, and we require  
\beq
R^3 \gg M
\eeq
while if the entries are random and independent of each other (and $N$), we have the requirement:
\beq
R^3 \gg \sqrt{N}M
\eeq
%The scalings with $M$ and $R$, here, are general, but there are assumptions which go into the scaling with $N$.  
%There are contributions from many types of flux to the mass matrix of the many moduli.  If all fluxes contribute to mass terms for all moduli,
%one might expect a matrix with random entries of order $\sqrt{N}$.  For such a random matrix, the eigenvalues would
%grow typically as $N^{3/2}$.  But one might well expect that the matrix would be more sparse, especially at large $R$.
%Conceivably there would be no growth of the eigenvalues of the mass matrix
%with $N$.  In a landscape, however, both $N$ and $M$ must be large,
In any case $R$ must be large, at least by powers of $M$, and quite possibly by powers of $N$ (this is certainly
the case with the assumptions of \cite{mcalister}).

Requiring sufficiently weak coupling (perturbativity) is potentially an even stronger constraint.
Couplings of complex structure moduli involving $n$ fields scale as
\beq
\lambda^{(n)} \sim M^2 R^{-4 -n}
\eeq
So if $R$ is the source of suppression of couplings, it appears one may require even larger $R$ as a function of $N$.
Alternatively, suppression may result from $g_s$ or from features of the string amplitudes themselves, as we will
discuss shortly.

The problem of fixing the radial moduli, especially without supersymmetry or approximate supersymmetry, is not well understood.  
Assuming that such large $R$ is required, it is natural to ask how typical this might be.
It is hard to make general statements about the non-supersymmetric case, but models for fixing radii with fluxes were put forward
in the large extra dimension scenario of \cite{add,sundruma,sundrumb,march,banksdinenelson}.  These involved playing off potentials due to fluxes with a bulk cosmological constant, curvature terms and possibly Casimir energies.  The typical scalings were designed to allow very large extra dimensions and vanishing four dimensional cosmological constant.  If terms of the first three types were of comparable importance, so as to allow cancellation of the c.c., then the typical radii satisfied
$R \sim M^{1/3}  $.  This is not good enough to give rise to many light fields.  Other scalings and tunings of parameters may
allow more rapid growth of $R$ with $M$, but it seems that large radius with many light fields is not likely to be typical.
In general, we might expect large $R$ to be associated with runaway behavior. An alternative approach is the exponentially large volume
scenario of \cite{quevedo}.   While perhaps not evidence for any particular phenomenon, the instability of typical vacua in large N models is consistent with the expectation that there should not be vast numbers of states at very large $R$.  We will return to this issue in section \ref{larger}.

The KKLT model\cite{kklt} provides a scenario, tied to approximate supersymmetry, in which a small fraction of states lie at large $R$.  (As a model of moduli fixing in a landscape, it is necessary that the actual number of such states still be substantial).  In this
model, there is a large hierarchy between the masses of the majority of the moduli (complex structure moduli) and the lightest (Kahler) moduli, controlled not only by $R$ but by the parameter $W_0$ (the expectation value of the superpotential).  
%Without approximate supersymmetry, arguments like those of \cite{add} lead to moduli masses of order the fundamental (string or cutoff) scale.
%In any case, field theory models with large numbers of fields provide an interesting arena for modeling a possible landscape, and it is interesting to ask what are the conditions in string theory that such a picture makes sense.
Specifically,
the radii are controlled by the Kahler modulus, $\rho \sim R^4$, which is itself of order:
\beq
\rho = -{1 \over b} \log (W_0)
\eeq
where $b$ is a number of order one.  So we require
\beq
W_0 \approx e^{-b \rho}.
\eeq
In the KKLT scenario, this {\it might} be of order the fraction of states with TeV scale supersymmetry
breaking, for example.  In any case, this might be a substantial set of states where large $R$ is self-consistent. 
But as we will argue, large radius is not likely to be a driving feature in the population of approximately supersymmetric states,
and not a critical issue in stability. 

An alternative picture, also with approximate supersymmetry {\it and} exponentially large volume was put forward in
ref. \cite{quevedo}.  The framework was closely tied to string theory.  A number of challenges for such a picture were discussed in \cite{dinefestucciaaxions},
but if successfully met, this would fall in the general class of approximately supersymmetric theories we have discussed,
and we would not expect any significant suppression of stability with the number of fields.

\subsection{Analogous Questions in Critical String Theory}

We have seen that unitarity/perturbativity place restrictions on behavior of couplings with $N$.  This might be enforced by
large radius or weak string coupling.  Alternatively, we have suggested a different possibility, that couplings might be ``diagonal", in the sense that, for example, the set of substantial quartic couplings might be sparse.  While it seems
challenging to address this question in an actual flux landscape, we can consider similar issues in critical string
theories with large numbers of fields.

We will consider supersymmetric theories, and suppose that we have a set of chiral fields, $\phi_i$, with Yukawa couplings 
(cubic superpotential couplings) among them.  First, let's review the
issue of unitarity/perturbativity in this context.  Suppose $X$ is a particular chiral multiplet, and that it couples
to a large number of chiral multiplets through couplings of the form $\gamma_i X^2 \phi_i$ (this will be 
relevant to the question of stability in the supersymmetric case shortly.
Consider the process $X + X \rightarrow X + X$ (where here $X$ denotes the scalar in the multiplet) at energies high compared to the masses of the $N$ $\phi_i$'s.  The potential includes a term
\beq
V(X) = \sum_{i=1}^N \vert \gamma_i \vert^2 \vert X \vert^4.
\label{vsusy}
\eeq
The cross section, then, behaves as
\beq
\sigma(X + X \rightarrow X + X) = {(\sum_{i=1}^N \vert \gamma_i \vert^2)^2 \over s}.
\eeq

With the assumption that all of the $\gamma_i$'s are chosen independently from the same distribution, the corresponding amplitude grows as $N^2$, and violates partial wave unitarity for large $N$ by a factor $N^2$; in other words, as one
increases $N$, it would be necessary to take the string coupling small as a power of $N$.  This would seem
surprising.

\subsection{A String Calculation}
\label{stringcalculation}

An example, which is relatively simple to analyze, is provided by compactification of the heterotic string on a Calabi-Yau space with large Euler number.  While there is not a systematic large $N$ computation (as is also true for most landscape constructions) we can still reasonably ask whether large numbers appear in perturbative computations.  This would imply, in a manner analogous to our landscape discussion, that a valid perturbation theory would require a large radius or very small $g$ (scaling as a power of $N$).  Here we can focus on the cubic terms in the
superpotential.  Consider, for example, the well-studied case of defined by the vanishing of a quintic polynomial in $CP^4$.There are $101$ $27$'s (corresponding to $101$ complex structure moduli).  At the level of cubic superpotential terms, $27^3$, there are of order $10^5$ independent couplings.  For special points in the moduli space, the actual number is much smaller, due to enhanced
symmetries.   For example, it is well known that there is a point with a $Z_5^4 \times S_5$ symmetry\cite{gsw}.  At this point there
are only a small number of non-vanishing couplings.  If one considers, for example couplings of the type $(a,a,b)$, with fixed $a$, there is typically only one field ($b$) which can combine to form an invariant, rather than $101$.  But at generic points in the moduli space, the symmetry is broken and one would expect all couplings to be non-vanishing.

On the other hand, there is a simple way to bound certain combinations of couplings, which we will see is relevant to this problem.  This arises from considerations of two dimensional conformal field theory (analogous considerations may be applicable in the framework of four dimensional theories, e.g. \cite{bootstrap}).   In such theories, coupling constants can be extracted
from correlation functions of {\it two dimensional} fields.  Features of such correlation functions, even for rather
complicated compactifications of the theory, can readily be extracted, particularly for large compactification radius,
where the two-dimensional, ``world-sheet" theory which governs the two dimensional dynamics, is weakly coupled.

Consider a four point function of vertex operators:
\beq
G(z_1,z_2,z_3,z_4)_{ab}=\langle V_a(z_1) V_b(z_2) V_a(z_3) V_b(z_4) \rangle
\eeq
This correlator yields the $S$-matrix of the particles labeled by $a$ and $b$.
If the leading term in the operator product expansion of $V_a$ and $V_b$ is:
\beq
V_a(z_1) V_b(z_2) = c_{abc} {V_c \over \vert z_1 - z_2 \vert^2},
\eeq
the $c_{abc}$ (up to normalization) are the three point couplings of the particles $a$, $b$ and $c$.
If we take $z_1 \rightarrow z_2$, $z_3 \rightarrow z_4$,
\beq
G(z_1,z_2,z_3,z_4)_{ab} = {\sum_{c=1}^N c_{abc}^2 \over \vert z_1 -z_3 \vert^2 \vert z_1 - z_2 \vert^2 \vert z_3 - z_4 \vert^4 }
\eeq
So if we can estimate or bound the four point function, we can bound the couplings of the fields $a,b$ to other fields.
In particular, if we consider a Calabi-Yau space, for example, with very large $N$, growth of the sum with $N$ would
require comparable growth of the correlation function.  

To illustrate the application of this type of analysis, consider the heterotic string compactified on a Calabi-Yau space.  We 
consider this class of models because, as we will see, the computation of the correlators is particularly simple.
We focus on cubic terms in the superpotential, i.e. terms of the type $27^3$ (we will comment on other string constructions in a future publication, as well as elaborating some technical aspects of the analysis below).  
Let's consider a Green's function which receives a contribution from the cubic superpotential terms, involving two
fermions and two bosons.  For definiteness and because of its simplicity,  we work in the fermionic formulation for the gauge
degrees of freedom of the string theory, and in the R-NS formulation for the right moving fermions.
Then the spatial coordinates can be grouped as $y^i, y^{\bar i}, x^\mu$, where the $i$, $\bar i$ are complex indices for the six dimensional Kahler manifold, and $\mu$ are four dimensional Minkowski indices.  The left moving fermions are $\lambda^i, \lambda^{\bar i}, \lambda^a$, where the $a$'s are $O(10)$ indices.  Space-time spinor operators can be taken as $S^0_\alpha, S^i_\alpha,S^{\bar 0}_\alpha, S^{\bar i}_\alpha$, where $\alpha$ are four dimensional spinor indices.  The $0$ and $\bar 0$ indices correspond to the covariantly constant spinor.  At large radius, the theory is nearly free, and these operators reduce to their free field forms.  Bosonizing the right moving fermions,
\beq
\psi^i = e^{i \phi_i}
\eeq
for the fermions with indices in the compact space,
whereas for the $\psi$'s with Minkowski indices
\beq
(\psi_1 + i \psi_2) = e^{i \xi_1}~~(\psi_3 + i \psi_4) = e^{i \xi_2}.
\eeq
Then
\beq
S^0 = c e^{{i \over 2} (\phi_1 + \phi_2 + \phi_3)}e^{{i \over 2} (\pm \xi_1 \pm \xi_2)}~~~S^{\bar 0} = c e^{-{i \over 2} (\phi_1 + \phi_2 + \phi_3)}e^{{i \over 2} (\pm \xi_1 \pm \xi_2)}
\eeq
where in the first case there are an even number of plus signs, the second an odd number.
The $S_i$'s are given by
\beq
S^i_\alpha = c e^{{i \over 2} (\phi_1 + \phi_2 -\phi_3)}e^{{i \over 2} (\pm \chi_1 \pm \chi_2)}~~~S^{\bar i}_\alpha = c e^{-{i \over 2} (\phi_1 + \phi_2 - \phi_3)}e^{{i \over 2} (\pm \xi_1 \pm \xi_2)}
\eeq
(this is $S^3, ~S^{\bar 3}$; other values of the index are obtained by changing the placement of the minus sign in the first exponent).
Decomposing the $27$ into representations of $O(10) \times U(1)$, 
\beq
27 = 16_{-1/2} + 10_1 + 1_{-2}
\eeq
the boson vertex operators for particles in the $1$ can be taken
to be :
\beq
V_B = \lambda^{\bar i} \lambda^{\bar j} \psi^k \chi^{(\alpha)}_{k \bar i \bar j}
\eeq
and its complex conjugate.  $\chi^{(\alpha)}_{k \bar i \bar j}$ is a harmonic (2,1) form.
It is related to the corresponding fluctuation in the metric, $\delta g_{ij}^{\alpha}$ through
\beq
\delta g^{(\alpha)}_{ij} = \chi_{~\bar k \bar l}^{\bar m {(\alpha)}} \Omega^{\bar k \bar l \bar n} g_{j \bar m}g_{i \bar n}.
\eeq
where $\Omega_{ijk}$ is the covariantly constant three form.
We will write the fermion vertex operator, for particles in the $10$ representation, as:
\beq
V_F = \epsilon_\alpha^i\lambda^a \lambda^i S^j_\alpha \delta g^{(\alpha)}_{ij}
\eeq
(it is also necessary to include suitable superconformal ghost fields in these vertex operators).

We consider the scattering of one fermion corresponding to the $(\alpha)$'th $(2,1)$ form with a scalar corresponding to the $\beta$'th $(2,1)$ form, to produce the same fermion and boson (elastic scattering).  This arises through the cubic terms in the superpotential,
\beq
\gamma_{\alpha \beta \gamma} \Phi^\alpha \Phi^\beta \Phi^\gamma.
\eeq 

In a sigma model description, at large $R$, the singular parts of the Green's function are readily evaluated, as the correlation functions of the  $\phi_i$'s and $\chi_i$'s (as well as the ghosts, which we have suppressed) are those of free fields.  The usual $\delta$ function for energy and momentum conservation in the compactified directions is replaced by an integral over the compact dimensions.
The coefficient of
\beq
{1 \over \vert z_1 -z_3 \vert^4 \vert z_1 - z_2 \vert^2 \vert z_3 - z_4 \vert^2 }
\eeq
is
\beq
{\cal A} = \int d^6 y  \delta g^{(\alpha)}_{ij} \delta g^{(\beta)}_{kl} \left ( \delta g^{*{(\alpha)}ij} \delta g^{*{(\beta)}kl} - \delta g^{*{(\alpha)}ik} \delta g^{*{(\beta)}jl} \right )
\eeq
The $\delta g$'s are normalized to unity.  As a result, the integral does not show growth with the number of complex structure moduli ($h_{2,1}$ of the manifold) unless the $\delta g$'s are singular.  On the other hand
\beq
{\cal A} = \sum_k \vert \gamma_{ijk} \vert^2,
\eeq
which is now of order $1$ rather than of order $N$.

This argument might fail if $\cal A$ were enhanced with $N$.  This in turn requires that the integral above be singular.  Taking the $\delta g$'s to be normalized, we can bound ${\cal A}$ by ${\rm max}~ \vert \delta g \vert^2$.  So we would require that the maximum be of order some power of $N$.  At the highly symmetric point in the moduli space studied in \cite{gsw}, there is no such enhancement, and it would be surprising if it arose as we turned out various complex structure moduli.  

Another argument for suppression comes from thinking about the mirror manifold\footnote{I thank S. Kachru for help in refining this argument}.  For the mirror, the Yukawa's we compute here are just intersection numbers.  These are expected to be sparse.
Applications of this type of analysis to other string theories and types of compactifications, as well as further investigations of possible loopholes, will be considered in a future publication.

%\section{Fluctuations of Masses}

\section{The Approximately Supersymmetric Case}
\label{approximatesusy}

For many years, supersymmetry has been viewed as a promising framework in which to resolve the hierarchy
problem.
In a landscape framework, it is not clear whether supersymmetry is particular important for questions of naturalness; it is simply
possible that there are overwhelmingly more non-supersymmetric than supersymmetric states\cite{douglasnonsusy,susskindnonsusy}.  Indeed, one might think
that supersymmetry is exceptional among possible states in a landscape.  The issue of stability, however, suggests a special role
for supersymmetry.  At least naively, one would expect that approximate supersymmetry would greatly increase
the chances for stability, as exact supersymmetry guarantees  exact stability.  So given the possibly extreme suppression of stability we have seen without supersymmetry,
it is conceivable that supersymmetric states are more common than non-supersymmetric states.

\subsection{Roads to Approximate Supersymmetry in Landscape Models}

Approximate supersymmetry has been modeled in landscapes in various ways.  Douglas and Denef\cite{douglasdenef}, starting
with an underlying supergravity theory, and making assumptions about distributions of parameters, argued that approximate
supersymmetry was highly suppressed.  This was shown to follow\cite{dinesun} from the assumption that such theories are
described, at low energies, by a single light chiral field, $Z$, with superpotential parameters uniformly distributed as complex numbers,
and Kahler potential parameters as real numbers.  This suppression, as the supersymmetry breaking scale, $F$, to the sixth power,
was seen as an argument that such theories do not resolve the hierarchy problem in a natural way.  Indeed, in this branch
of the landscape, low energy supersymmetry, when it occurs, is an accident, arising from tuning fluxes to obtain
a low energy theory with approximate supersymmetry.  We will refer to models of this type as models with {\it tuned supersymmetry}.

Within theories with $N$ fields, on such a branch, one field is by assumption far lighter than the others.  The ``price" of this tuning
is  part of the $F^6$ suppression.  {\it With this assumption}, it would seem surprising that there would be extreme suppression
beyond that of \cite{douglasdenef} from the requirement of stability.  Indeed, as there is
no requirement of an additional light chiral field, one would expect that all of the other
moduli are typically heavy.    For example, if almost all of the  moduli have masses suppressed by powers of $R$,
while the lightest modulus is far lighter, then we would not expect a dramatic, qualitative change as the radius becomes
of order one.  Indeed, we will argue in the next subsection that there is likely, at most, an order one suppression

A more plausible setting for supersymmetry arises if the breaking is dynamical\cite{branches,dinegorbatovthomas}.  In this case, while very low energy
breaking of supersymmetry is not necessarily favored, it is typically not disfavored.  Such theories exhibit an exponential
gap between the scales of supersymmetry breaking and other states.  Certainly the presence of
$N$ relatively heavy ``moduli" should not have a strong bearing on questions of stability.  We will understand these
points in the next section.

\subsection{Classical Stability with Approximate Supersymmetry}

Refs. \cite{mcalister,Bachlechner:2014rqa}, studied theories with approximate supersymmetry (in the sense of tuned supersymmetry, defined above) and $N$ massive fields.  Here ``massive" means heavy compared to any scale of supersymmetry breaking, but light compared to the string (or other fundamental) scale.  They argued that one expects an exponential suppression of classical stability, i.e. only an order $e^{-cN}$ fraction of states are classically stable.

At first sight,  this result is puzzling.  In these cases, one would expect that there is a scale with approximate supersymmetry with only one chiral field.  There would then be only a small number of important parameters in the effective action.  While there would be some constraints on these parameters, it is surprising that the stability of a few light scalar fields is exponentially sensitivity to the number of heavy fields (indeed, any significant suppression with $N$).

Calling the Goldstino supermultiplet $X$,
with the scalings with $N$ we have described in eqn. \ref{nscalings},
we can take its superpotential to be $fX$, and its Kahler potential $X^\dagger X$, up to terms suppressed by powers of $N$.
Requiring (nearly) vanishing c.c., would
yield exactly the Polonyi model, for which the scalar fields have positive mass-squared.  Negative mass-squared would
seem not only unlikely but impossible.

It is easy, however, to identify where the sensitivity to the number of heavy fields arises.  Assume that, in addition to $X$,
there are $N$ heavy fields, $\Phi_i$.  Suppose the superpotential for $X$ and $\Phi_i$ is:
\beq
W = X f + \sum_{i=1}^N \left ({1 \over 2} m_i^2 \Phi_i \Phi_i + \gamma_{i} X^2 \Phi_i \right ).
\eeq
Integrating out the massive fields generates a correction to the Kahler potential:
\beq
\delta K = \sum_{i=1}^N{\vert \gamma_i^2\vert  \vert X^\dagger X \vert^2 \over m_i^2}.
\label{kahlercorrection}
\eeq
This leads to a correction to the mass of $X$.
One can calculate this easily  by looking directly at the potential and integrating out the $\Phi_i$ fields by solving for the minimum of their
potential as a function of $X$.  The result is:
$$m_X^2 = -\sum_{i=1}^N{ \vert \gamma_i \vert^2 \over 4 m_i^2}$$
in agreement with formulas in \cite{mcallister}.

It is important to understand in this expression how the various couplings scale with $N$.  If, for example, as suggested in
\cite{mcalister,Bachlechner:2014rqa}, the $\gamma$'s scale as $1/\sqrt{N}$, given that there are $N$ positive terms in the sum,
whether the $X$ mass is typically tachyonic depends on how $1/m_i^2$
behaves.  The mass matrix is of the form $m^\dagger m$ (a {\it Wishart} matrix, see, e.g. \cite{mcalister}).  These authors assumed that
the the entries in $m$ are of order $1/\sqrt{N}$  As a result, the typical masses are of order $1$, but
there are frequent fluctuations to low eigenvalues, and it is rare that the sum is
small; indeed, a numerical study indicates that the sum is typically of order $N$.  With these assumptions,
only exponentially rare downward fluctuations will lead to stability.

We have advocated a different scaling above, one which is compatible with perturbativity (and also with diagonal
couplings).  With these scalings, there is an additional $1/N$ (the scaling of masses is the same, but the
couplings at the vertices scale as $1/N$).  As a result, the sum of terms is typically of 
order one, i.e. comparable to the leading Polonyi term.   (Here, the couplings give $1/N^2$, the inverse mass squared
factor gives a factor of order $N$, and the overall number of terms gives an additional $N$).  So these affects don't grow with $N$. 
As a result, the chances of stability are of order one.  Note that this extra contribution to the Kahler potential is soft, and so
consistent with our scalings. 
%There may be additional sources of mass for $X$, but this term, with it's particular $N$ dependence, is special.  
%This is competitive with the lowest order contribution.  But unless there is some enhancement with $N$, one expects at most
%an order one suppression, not an effect exponential in $N$.

It should be noted that there might be additional suppression of masses.  In the string context, this could arise due to large $R$, as we have discussed.
In this case, however, there is additional, compensating dependence of couplings with $R$, and as a result one does not expect further difficulties
with stability.

%It should be noted that this order one result is consistent with our requirements for a valid perturbation expansion. because the term is soft, diagrams with this coupling are quadratically divergent, with a cutoff $1/N^2$.

This does not change the fact that tuned supersymmetry is expected to be rare in a landscape.
On the other hand, heavy fields do not pose any issue for dynamical supersymmetry breaking.
In such theories, the mass of partners of the Goldstino (if such can be identified) are typically of order
the scale of supersymmetry breaking, rather than the gravitino mass.  So the contributions to the Kahler
potential from integrating out very massive fields are totally inconsequential.
.

\section{Landscape Statistics at Large $R$}
\label{larger}

We can ask whether the various speculations about stability are compatible with the picture of large numbers of fields
arising at large $R$ in compactified string theories.
The presumption of an exponentially large number of states at large $R$, we have already argued,
is not terribly plausible.  Another way to see the puzzle this raises is to consider how the distribution of states might behave with $R$.
If for large $R$, there is an exponentially large number of states with some parameter $N$, do the majority of these simply
disappear -- along with the parameter $N$ -- as $R$ decreases?  This would seem particularly troubling in the supersymmetric
case.  If not, what would play the role of the parameter $N$ at small $R$?

The high level of instability in the non-supersymmetric case would seem to avoid this paradox.  The exponential suppression
could be take to indicate that there is typically at most of order one metastable state per flux choice.  Assuming this is the case
(and that the number is not significantly smaller than one), the question becomes:  what is the rate for tunneling between states with different flux.  In other words, quantum stability becomes the critical issue.

In the case of tuned supersymmetry,  one does not expect particularly strong dependence on $R$ in the counting of states, and this is compatible
with our arguments that there is no $N$ dependent suppression of stability.  Such a picture, indeed, avoids any
puzzles about discontinuous changes in the population of states as $R$ is varied; the same level of tuning is required in any
case.  With dynamical supersymmetry breaking, this issue does not arise.

\section{Quantum Stability}
\label{quantumstability}

In the event that a state is classically stable, it is necessary to ask about quantum stability.  This question takes on particular
significance in a landscape context.  In the flux vaccum picture, any would-be nearly Minkowski vacuum will be surrounded by vast numbers of states with negative cosmological constant.  With large numbers of fields, there will be many possible tunneling trajectories.  Every one of these must be significantly suppressed.

This problem has been discussed elsewhere, but the observations of \cite{mcalister} and this paper permit one to make
further statements about stability against tunneling.  In \cite{greeneweinberg}, models with large numbers of fields were studied (without
supersymmetry).
Starting with a would-be classical ground state, the authors Taylor-expanded the potential, and assumed that the coefficients of the expansion
were uniformly distributed random variables.  In \cite{dinepaban}, the results of \cite{greeneweinberg} were understood in a simple
way:  roughly speaking, the most probable tunneling trajectory is a straight line in field space in the direction of the lightest field. 
Suppressing tunneling then requires that all masses be larger than some minimum value.  More precisely, the ratio of
$\mu^2 \over \gamma^2$, where $\gamma$ is the cubic coupling in the direction of the field of mass $\mu$, for every
choice of field, must be larger than some minimum value.  With the assumption of uniform distribution of couplings,
with the intervals considered in \cite{greeneweinberg}, this leads to exponentially small probability for bounce actions all greater
than some minimum, $B_0$.  

This analysis must be rethought in light of these observations.  First, the assumption that {\it all} couplings are chosen from the same, $N$-independent random distributions, runs afoul of unitarity and perturbativity.  If we assume, instead,
$N$-dependent distributions, with couplings falling as powers of $N$, then qualitatively different results than those
of \cite{greeneweinberg, dinepaban} may emerge.  In particular, if the cubic couplings fall off more 
rapidly than $\mu^2$ with $N$, then bounce actions may be enhanced and tunneling suppressed.
Indeed, the scalings above would suggest that the typical bounce action grows rapidly with $N$.
Alternatively, if couplings and masses are diagonal in the same basis, the analysis of \cite{dinepaban}
implies significant further suppression of the tunneling.  Indeed, it would seem surprising if classical stability were highly suppressed,
but most classically stable vacua were highly metastable.  Other than arguing that this is counterintuitive, however,
we don't have a sharp argument for this type of alignment.

In flux landscapes, for $R \sim 1$, with an order one number of states per choice of flux, without supersymmetry,
ref. \cite{dinefestucciamorisse} argued that tunneling rates were fast, with bounce actions behaving
as inverse powers of flux.

With approximate supersymmetry, as stressed in \cite{dinefestucciamorisse}, the situation is different.  In field directions
with curvature much greater than $m_{3/2}$, tunneling is highly suppressed, either vanishing or
scaling as
\beq
\Gamma \propto e^{-M_p^2/m_{3/2}^2}.
\eeq
Assuming that there are only
a small number of fields with mass of order $m_{3/2}$ only a few tunneling amplitudes need to be
suppressed ((roughly corresponding to tunneling
in directions connected with the goldstino multiplet), so one does not expect to pay an exponential price.  These statements hold not only for dynamical
supersymmetry breaking, where classical stability is not a severe constraint, but also to the case where low energy supersymmetry
is the result of tuning of parameters (fluxes).

\section{Conclusions}
\label{nonsusy}

The existence of a landscape remains a subject of conjecture, based mainly on considerations of models of multiple
fields and on the many candidate choices of flux in string theories.  In this context, questions of stability, classical and
quantum, can contribute both to assessment of the plausibility of the basic picture, and to understanding what features a putative
landscape might exhibit.  In the second category, from the beginning there has been a question:  does supersymmetry
play any role?  Underlying this question is a view that supersymmetric states might be exceptional, a view which finds support
in the work
of Douglas and Denef\cite{douglasdenef}, indicating that small supersymmetry breaking is rare.

The considerations of stability which we have reviewed and extended here provide insight into both classes of questions.
The work of \cite{mcalister} (and earlier \cite{easther}) indicates that, without supersymmetry, classical stability
might be extraordinarily rare.  Within flux vacua, this might mean that for a given choice of flux, there is at most one stable
state (with runaway likely being typical).  We have indicated that some of the assumptions which go into this analysis
{\it might} not be correct; critical string theories give some evidence that couplings may be highly correlated, and
perturbativity requires scalings of couplings similar to what one would expect were they diagonal
in some basis.  At the same time, we have argued that large numbers of fields (moduli) in string theory compactifications
with fluxes require very large radius, and that this is likely to be very rare.  But our remarks on this question
are only conjectural.

Without supersymmetry,
quantum stability potentially provides further constraints, but these 
are also dependent on assumptions about distributions of couplings.  If we require perturbativity but assume couplings are 
uncorrelated, the additional suppression could be quite mild; but further exponential suppression is also a possibility.

Supersymmetry (i.e. supersymmetry broken at scales well below the fundamental scale) does appear to occupy a priviedged 
place in any would-be landscape.  General principles suggest that stability should be typical of states found
in any sort of analysis.  We have distinguished tuned supersymmetry breaking and dynamical breaking.
In the former case, we have understood a possible source of instability noted in \cite{mcalister}, and that this
is unlikely to yield further appreciable suppression.  We have also noted that for dynamical supersymmetry
breaking, it is irrelevant.

We have discussed the question of large numbers of fields {\it within} flux landscapes.  We have noted that such a picture
is only valid when the compactification radius is large, scaling with a power of the flux.  The power itself depends on
assumptions about the nature of the underlying effective action.  From the start, the existence of exponentially larger
numbers of states at {\it large} radius relative to those at smaller scale is not terribly plausible.   We have put forward a consistency condition that the
spectrum of {\it stable} states should not depend critically on $R$.  For the non-supersymmetric states, this is consistent
with a vast suppression.  For the supersymmetric case, our discussion suggests that large $R$ is not special.

If there is of order one non-supersymmetric state, classically, for typical choices of fluxes, absent supersymmetry there remains the question
of quantum stability.  Arguments that instability is typical, in this case, were put forward in \cite{dinefestucciamorisse}.

There are a few principled things one can say about the case of approximate supersymmetry.
\begin{enumerate}
\item  With zero cosmological constant and exact supersymmetry states are automatically stable.% or nearly so.  
\item  With approximate supersymmetry and small cosmological constant, there are only a small number of relevant light fields (i.e. we do not expect scaling with $N$), so one does not expect exponential suppression of either classical or quantum stability with $N$.
\item  Supersymmetry and approximate supersymmetry also guarantee stability in cases where, for a fixed choice of flux, there are only small numbers of states.
\end{enumerate}
%Large numbers of fields do not provide a particularly interesting arena for modeling a landscape of supersymmetric states.

All of this suggests that supersymmetry, at a scale significantly below the fundamental scale, is likely to be important in a landscape.  Stability does not, by itself, argue for TeV scale supersymmetry.

\vskip .2cm
\noindent
%{\bf Acknowledgements:} 
\noindent
{\bf Acknowledgements:}  We thank Raphael Bousso, Shamit Kachru, Liam McAlister,  Yasnori Nomura, Steve Shenker and Sonya Paban for sharing their insights into these questions.  %This work was supported in part by the U.S. Department of Energy grant number DE-FG02-04ER41286. 
This work was supported by the U.S. Department of Energy grant number DE-SC0010107.

%\bibliography{unitarity_constraints_refs}{}
%\bibliographystyle{utphys}
\bibliography{jhep_revised_classical_and_quantum_stability.bbl}{}

%\bibliographystyle{unsrt}
%\bibliographystyle{JHEP}
%\bibliography{Biblio}

\end{document}